\documentclass[12pt]{article}
\usepackage{amssymb}
\usepackage{amsmath}

\newcommand{\la}{\lambda}

\newcommand{\prt}{\partial}

\begin{document}

\title{On generating functions in the AKNS hierarchy}
\author{A.M. Kamchatnov and M.V. Pavlov\\
Institute of Spectroscopy, Russian Academy of Sciences,\\
Troitsk, Moscow Region, 142190, Russia }

\maketitle

\begin{abstract}
It is shown that the self-induced transparency equations can be interpreted
as a generating function for as positive so negative flows in the AKNS
hierarchy. Mutual commutativity of these flows leads to other hierarchies
of integrable equations. In particular, it is shown that stimulated
Raman scattering equations generate the hierarchy of flows which include the
Heisenberg model equations.
This observation reveals some new relationships between known
integrable equations and permits one to construct their new physically important
combinations. Reductions of the AKNS hierarchy to ones with complex conjugate and real
dependent variables are also discussed
and the corresponding generating functions of positive and negative flows are found.
Generating function of Whitham modulation equations in the AKNS hierarchy is obtained.
\end{abstract}



It is well known that many physically important integrable partial differential equations
belong to the AKNS hierarchy \cite{AKNS}. Up to now most attention was paid to its
positive flows, where both $2\times 2$ matrices $\mathbb{U}$ and $\mathbb{V}$
have matrix elements polynomial in the spectral parameter $\la$ what leads to
the recursive structure of the hierarchy so that subsequent flows are connected
by the recursion operator (see, e.g. \cite{Newell}). However, negative flows in the AKNS hierarchy
have not been considered systematically enough, though the sine-Gordon equation and
its connection with the mKdV hierarchy has been a recurrent theme in the
soliton literature (see, e.g., \cite{Chodos}--\cite{FMG}). Another example of well-studied
negative flow is provided by the self-induced transparency (SIT) equations \cite{Lamb,AKN}
which appeared also in different forms and contexts as Pohlmeyer-Lund-Regge
equations \cite{Pohlmeyer}--\cite{Lund}. Their role as a symmetry flow in the
AKNS hierarchy has been recently considered in \cite{AFGZ} in framework of
loop algebra approach to the integrable equations. Here we remark that SIT equations
can be interpreted as a generating function of positive and negative flows in the
AKNS hierarchy. This leads to simple method of obtaining the integrable equations of the
hierarchy and corresponding Lax pairs. Generalization of this point of view leads to new useful
connections between integrable equations belonging to as positive so negative
flows in the AKNS hierarchy.

The AKNS hierarchy \cite{AKNS} is based on the
Zakharov and Shabat \cite{ZSh} spectral problem
\begin{equation}\label{eq1}
\left(
\begin{array}{c}
\psi_1 \\ \psi_2
\end{array}
\right)_x=
\left(
\begin{array}{cc}
\la & q \\
r & -\la
\end{array}\right)
\left( \begin{array}{c}
\psi_1 \\ \psi_2
\end{array}\right).
\end{equation}
Let us consider the first negative flow with the pole at $\la=\zeta$ in the
complex plane of the spectral parameter $\la$:
\begin{equation}\label{eq2}
\left(
\begin{array}{c}
\psi_1 \\ \psi_2
\end{array}
\right)_{\tau}=\frac1{\la-\zeta}
\left(
\begin{array}{cc}
a & b \\
c & -a
\end{array}\right)
\left( \begin{array}{c}
\psi_1 \\ \psi_2
\end{array}\right).
\end{equation}
The compatibility condition of systems (\ref{eq1}) and (\ref{eq2})
yields equations
\begin{equation}\label{eq3}
    \begin{array}{l}
    a_x=cq-br,\quad b_x-2\zeta b=-2aq,\quad c_x+2\zeta c=2ar,\\
    q_{\tau}=-2b,\qquad r_{\tau}=2c,
    \end{array}
\end{equation}
which can be reduced to well-known SIT equations with $\zeta$ playing the role
of the ``detuning'' parameter \cite{Lamb,AKN}. Then, introduction of a hierarchy of
times $t_n$ labelled by inverse powers of $\zeta$,
\begin{equation}\label{eq4}
    \frac{\prt}{\prt\tau}=\sum_{n=0}^{\infty}{\zeta^{-n}}\frac{\prt}{\prt t_n},
\end{equation}
and of expansions of $a$, $b$, $c$ in inverse powers of $\zeta$,
\begin{equation}\label{eq5}
    a=\sum_{n=0}^{\infty}\zeta^{-n}a_n,\quad
    b=\sum_{n=0}^{\infty}\zeta^{-n}b_n,\quad
    c=\sum_{n=0}^{\infty}\zeta^{-n}c_n,
\end{equation}
together with geometric series expansion of $1/(\la-\zeta)$ in powers of $\la/\zeta$
leads at once to well-known recurrence relations for the positive AKNS hierarchy \cite{Newell},
\begin{equation}\label{eq6}
\begin{array}{ll}
    \sum_{n=0}^{\infty}{\zeta^{-n}}\left(
    \begin{array}{c}
    \psi_1\\ \psi_2
    \end{array}\right)_{t_n}&=\sum_{n=0}^{\infty}{\zeta^{-n}}\left(
    \begin{array}{cc}
    A_n & B_n\\ C_n & -A_n
    \end{array}\right)\left(
    \begin{array}{c}
    \psi_1\\ \psi_2
    \end{array}\right)\\ & \equiv
    \sum_{n=0}^{\infty}{\zeta^{-n}}\mathbb{V}_n\left(
    \begin{array}{c}
    \psi_1 \\ \psi_2
    \end{array}\right),
    \end{array}
\end{equation}
where
\begin{equation}\label{eq7}
\begin{array}{l}
    A_0=-a_0,\quad B_0=-b_0,\quad C_0=-c_0;\\
    A_n=\la A_{n-1}-a_n,\quad B_n=\la B_{n-1}-b_n,\quad C_n=\la C_{n-1}-c_n,
    \end{array}
\end{equation}
and $a_n$, $b_n$, $c_n$ can be found by substitution of (\ref{eq4}) and (\ref{eq5})
into (\ref{eq3}) and equating coefficients in terms with the same powers of $\zeta^{-1}$.
In this way with the choice of constants $a_0=1,$ $a_1=0$ we arrive at well-known
equations of the positive AKNS hierarchy
\begin{equation}\label{eq8}
    \begin{array}{l}
    q_{t_1}=-2q,\qquad r_{t_1}=2r;\\
    q_{t_2}=-q_x,\qquad r_{t_2}=-r_x;\\
    q_{t_3}=\frac12(-q_{xx}+2q^2r),\quad r_{t_3}=\frac12(r_{xx}-2r^2q);\\
    q_{t_4}=\frac14(-q_{xxx}+6qrq_x),\quad r_{t_4}=\frac14(-r_{xxx}+6qrr_x);\ldots ,
    \end{array}
\end{equation}
and corresponding matrices $\mathbb{V}_n$ governing the $t_n$ evolution of $\psi$ function
(see (\ref{eq6})).

In a similar way we obtain negative flows of the AKNS hierarchy, if we expand (\ref{eq2})
and (\ref{eq3}) in positive powers of $\zeta$,
\begin{equation}\label{eq9}
\begin{array}{l}
    \frac{\prt}{\prt\tau}=\sum_{n=1}^{\infty}{\zeta^n}\frac{\prt}{\prt t_{-n}},\\
    a=\sum_{n=1}^{\infty}\zeta^{n}a_{-n},\quad
    b=\sum_{n=1}^{\infty}\zeta^{n}b_{-n},\quad
    c=\sum_{n=1}^{\infty}\zeta^{n}c_{-n},
    \end{array}
\end{equation}
that is
\begin{equation}\label{eq11}
\begin{array}{ll}
    \sum_{n=1}^{\infty}{\zeta^n}\left(
    \begin{array}{c}
    \psi_1\\ \psi_2
    \end{array}\right)_{t_{-n}}&=\sum_{n=1}^{\infty}{\zeta^n}\left(
    \begin{array}{cc}
    A_{-n} & B_{-n}\\ C_{-n} & -A_{-n}
    \end{array}\right)\left(
    \begin{array}{c}
    \psi_1\\ \psi_2
    \end{array}\right)\\ & \equiv
    \sum_{n=0}^{\infty}{\zeta^n}\mathbb{V}_{-n}\left(
    \begin{array}{c}
    \psi_1 \\ \psi_2
    \end{array}\right),
    \end{array}
\end{equation}
where
\begin{equation}\label{eq12}
    \begin{array}{l}
    A_{-1}=a_{-1}/\la,\quad B_{-1}=b_{-1}/\la,\quad C_{-1}=c_{-1}/\la;\\
    A_{-n}=(A_{-n+1}+a_{-n})/\la,\quad B_{-n}=(B_{-n+1}+b_{-n})/\la,\\
    C_{-n}=(C_{-n+1}+c_{-n})/\la,
    \end{array}
\end{equation}
so that we arrive at equations of negative flows in the AKNS hierarchy,
\begin{equation}\label{eq13}
    (qr)_{t_{-1}}=2a_{-1,x},\quad q_{xt_{-1}}=4qa_{-1},\quad
    r_{xt_{-1}}=4ra_{-1};
\end{equation}
\begin{equation}\label{eq14}
\begin{array}{l}
(qr)_{t_{-2}}=2a_{-2,x},\quad q_{xt_{-2}}r-qr_{xt_{-2}}=4a_{1,x},\\
(q_{xt_{-2}}-4qa_{-2})_x=8qa_{-1},\quad
(-r_{xt_{-2}}+4ra_{-2})_x=8ra_{-1}; \ldots .
\end{array}
\end{equation}

It is known that all these flows commute with each other. Therefore we
can choose any two flows and find the corresponding compatibility condition
in terms of two relevant independent variables $t_m$ and $t_n$. Formally,
this corresponds to exclusion of $x$ variable from equations for $(x,t_m)$ and
$(x,t_n)$ flows by means of redefinition of the dependent variables \cite{GGH}.
This is possible as far as no dependent on $x$ reduction is imposed
on the $t_m$ and $t_n$ flows. However, this procedure is quite tedious and
in practice we can avoid it by working directly with the linear systems
\begin{equation}\label{eq14a}
    \psi_{t_m}=\mathbb{V}_m\psi, \qquad \psi_{t_n}=\mathbb{V}_n\psi.
\end{equation}
Then we obtain equations expressed in terms of new variables given by the
coefficients of $\la^k$ in the matrix elements of $\mathbb{V}_m$ and $\mathbb{V}_n$.
For example, let us consider flows arising
from commutation of the system
\begin{equation}\label{eq15}
\left(
\begin{array}{c}
\psi_1 \\ \psi_2
\end{array}
\right)_{t_{-1}}=\frac1{\la}
\left(
\begin{array}{cc}
a_{-1} & b_{-1} \\
c_{-1} & -a_{-1}
\end{array}\right)
\left( \begin{array}{c}
\psi_1 \\ \psi_2
\end{array}\right),
\end{equation}
with other systems corresponding to negative flows $t_{-n}$. In fact, with the use
of the above method, we can find the generating function of these flows.
To this end, we derive the compatibility conditions of (\ref{eq15}) and (\ref{eq2}),
\begin{equation}\label{eq17}
    \begin{array}{c}
    a_{-1,\tau}={a}_{t_{-1}},\qquad a_{-1,\tau}=\frac1{\zeta}(b_{-1}{c}-{b}c_{-1}),\\
    b_{-1,\tau}={b}_{t_{-1}},\qquad b_{-1,\tau}=\frac2{\zeta}(a_{-1}{b}-{a}b_{-1}),\\
    c_{-1,\tau}={c}_{t_{-1}},\qquad c_{-1,\tau}=\frac2{\zeta}(c_{-1}{a}-{c}a_{-1}).
    \end{array}
\end{equation}
These equations lead at once to the constraints
\begin{equation}\label{eq18}
    a_{-1}^2+b_{-1}c_{-1}=f_1(t_{-1}),\qquad {a}^2+{b}{c}=f_2(\tau)
\end{equation}
and after appropriate definition of dependent variables they reduce to a particular
case of stimulated Raman scattering equations \cite{Kaup,Steudel}.
Now, expansions of (\ref{eq2}) and (\ref{eq17}) with the use of (\ref{eq9})
yield the negative flows in terms of independent variables $t_{-1}$ and $t_{-n}$.
In particular, the second negative flow takes the form of the Heisenberg
model equations
\begin{equation}\label{eq20}
\begin{array}{l}
    a_{-1,t_{-2}}=\frac14(b_{-1}c_{-1,t_{-1}t_{-1}}-c_{-1}b_{-1,t_{-1}t_{-1}}),\\
    b_{-1,t_{-2}}=\frac12(a_{-1}b_{-1,t_{-1}t_{-1}}-b_{-1}a_{-1,t_{-1}t_{-1}}),\\
    c_{-1,t_{-2}}=\frac12(a_{-1}c_{-1,t_{-1}t_{-1}}-c_{-1}a_{-1,t_{-1}t_{-1}}),
    \end{array}
\end{equation}
and from (\ref{eq2}) we obtain the system
\begin{equation}\label{eq21}
\left(
\begin{array}{l}
\psi_1 \\ \psi_2
\end{array}
\right)_{t_{-2}}=
\left(
\begin{array}{cc}
A_{-2} &
B_{-2} \\
C_{-2} &
-A_{-2}
\end{array}\right)
\left( \begin{array}{c}
\psi_1 \\ \psi_2
\end{array}\right)
\end{equation}
\begin{equation*}\label{eq21a}
    \begin{array}{l}
    A_{-2}=\frac{a_{-1}}{\la^2}+\frac1{4\la}(b_{-1}c_{-1,t_{-1}}-c_{-1}b_{-1,t_{-1}}),\\
    B_{-2}=\frac{b_{-1}}{\la^2}+\frac1{2\la}(a_{-1}b_{-1,t_{-1}}-b_{-1}a_{-1,t_{-1}}) ,\\
    C_{-2}=\frac{c_{-1}}{\la^2}-\frac1{2\la}(a_{-1}c_{-1,t_{-1}}-c_{-1}a_{-1,t_{-1}})
    \end{array}
\end{equation*}
whose compatibility condition with (\ref{eq15}) yields equations (\ref{eq20}). In fact,
these two systems have the known form \cite{Takhtajan} with the spectral parameter
$\la$ replaced by $1/\la$.

From this point of view, the Heisenberg equations belong to the negative part of the
AKNS hierarchy. Hence, we can take a linear combination of systems (\ref{eq1}) and
(\ref{eq21}) commuting with (\ref{eq15}) to obtain the pair (with evident change of the notation)
\begin{equation}\label{eq22}
\begin{array}{l}
\left(
\begin{array}{c}
\psi_1 \\ \psi_2
\end{array}
\right)_{x}=\frac1{\la}
\left(
\begin{array}{cc}
a & b \\
c & -a
\end{array}\right)
\left( \begin{array}{c}
\psi_1 \\ \psi_2
\end{array}\right),\\
\left(
\begin{array}{c}
\psi_1 \\ \psi_2
\end{array}
\right)_{t}=
\left(
\begin{array}{cc}
A & B  \\
C  & -A
\end{array}\right)
\left( \begin{array}{c}
\psi_1 \\ \psi_2
\end{array}\right)
\end{array}
\end{equation}
\begin{equation}\label{eq23}
\begin{array}{l}
    A=\alpha\la+\frac{\beta a}{\la^2}+\frac{\beta}{4\la}(bc_{x}-cb_{x}),\quad
B=\alpha q+\frac{\beta b}{\la^2}+\frac{\beta}{2\la}(ab_{x}-ba_{x}),\\
C=\alpha r+\frac{\beta c}{\la^2}-\frac{\beta}{2\la}(ac_{x}-ca_{x}),
\end{array}
\end{equation}
where $\alpha+\beta=1$.
The compatibility condition of systems (\ref{eq22}) yields the equations
\begin{equation}\label{eq24}
    \begin{array}{l}
    a_t=\frac{\beta}4(bc_{xx}-cb_{xx})-\alpha(rb-qc),\\
    b_t=\frac{\beta}2(ab_{xx}-ba_{xx})-2\alpha qa,\\
    c_t=-\frac{\beta}2(ac_{xx}-ca_{xx})+2\alpha ra,\\
    q_x=-2b, \qquad r_x=2c.
    \end{array}
\end{equation}
Remarkably enough, this system coincides with
the equations describing self-induced transparency of light propagation through
medium with spatial dispersion \cite{AR}. Apparently, the fact of complete
integrability of these physically important equations has not been noticed yet.

In physical applications the dependent variables are usually subject to some constraints.
For example, the variables $q$ and $r$ may be complex (anti)conjugated to one another,
\begin{equation}\label{eq1'}
    r=\pm q^*.
\end{equation}
It is easy to find that the generating function of equations in the reduced AKNS hierarchy
based on the Zakharov-Shabat spectral problem
\begin{equation}\label{eq2'}
\left(
\begin{array}{c}
\psi_1 \\ \psi_2
\end{array}
\right)_x=
\left(
\begin{array}{cc}
\la & q \\
\pm q^* & -\la
\end{array}\right)
\left( \begin{array}{c}
\psi_1 \\ \psi_2
\end{array}\right)
\end{equation}
is given by
\begin{equation}\label{eq3'}
\left(
\begin{array}{c}
\psi_1 \\ \psi_2
\end{array}
\right)_{\tau}=\frac1{\la-i\zeta}
\left(
\begin{array}{cc}
a & b \\
\mp b^* & -a
\end{array}\right)
\left( \begin{array}{c}
\psi_1 \\ \psi_2
\end{array}\right),
\end{equation}
where $\zeta$ is considered as a real parameter. Then the compatibility conditions
take the form
\begin{equation}\label{eq4'}
    q_{x\tau}=4aq+2i\zeta q_{\tau},\quad a_x=\pm\frac12(|q|^2)_{\tau},\quad b=-\frac12q_{\tau},
\end{equation}
and expansions (\ref{eq4},\ref{eq5}) and  (\ref{eq9}) give, respectively, positive flows,
\begin{equation}\label{eq5'}
\begin{array}{l}
    q_{t_1}-2iq=0,\quad q_{t_2}-q_x=0,\quad 2iq_{t_3}-q_{xx}\pm2|q|^2q=0,\\
    4q_{t_4}+q_{xxx}\pm 6|q|^2q_x=0,\ldots,
    \end{array}
\end{equation}
and negative flows
\begin{equation}\label{eq6'}
\begin{array}{l}
q_{xt_{-1}}=4a_{-1}q,\quad a_{-1,x}=\pm\frac12(|q|^2)_{t_{-1}};\\
(q_{xt_{-2}}-4a_{-2}q)_x=8ia_{-1}q,\quad a_{-2,x}=\pm\frac12(|q|^2)_{t_{-2}},\\
a_{-1,x}=\pm\frac{i}4(qq_{xt_{-2}}^*-q^*q_{xt_{-2}}); \ldots .
\end{array}
\end{equation}

It is known (and easy to see from the above equations) that only even positive flows
and odd negative flows are consistent with the reduction
\begin{equation}\label{eq7'}
    r=\pm q \qquad ({\rm or}\quad q^*=q).
\end{equation}
These flows correspond to the mKdV-sine-Gordon
hierarchy based on the spectral problem
\begin{equation}\label{eq25}
\left(
\begin{array}{c}
\psi_1 \\ \psi_2
\end{array}
\right)_x=
\left(
\begin{array}{cc}
\la & q\\
\pm q & -\la
\end{array}\right)
\left( \begin{array}{c}
\psi_1 \\ \psi_2
\end{array}\right).
\end{equation}
It is easy to show that the generating function of this hierarchy is given by
\begin{equation}\label{eq26}
\left(
\begin{array}{c}
\psi_1 \\ \psi_2
\end{array}
\right)_{\tau}=\frac1{\la^2-\zeta^2}
\left(
\begin{array}{cc}
\la a & \mp \la b+c\\
\la b\pm c & -\la a
\end{array}\right)
\left( \begin{array}{c}
\psi_1 \\ \psi_2
\end{array}\right).
\end{equation}
The compatibility conditions of these two systems have the form
\begin{equation}\label{eq27}
    a_x=2qb,\quad b_x=\pm 2(aq-c),\quad c_x=-\zeta^2q_{\tau},\quad
    q_{\tau}=\pm 2b.
\end{equation}
Now positive flows correspond to expansions of (\ref{eq26}) and (\ref{eq27})
in powers of $\zeta^{-2}$:
\begin{equation}\label{eq28}
\begin{array}{l}
    \frac{\prt}{\prt\tau}=\sum_{n=0}^{\infty}{\zeta^{-2n}}\frac{\prt}{\prt t_{2n}},\\
    {a}=\sum_{n=0}^{\infty}\zeta^{-2n}a_{2n},\quad
    {b}=\sum_{n=0}^{\infty}\zeta^{-2n}b_{2n},\quad
    {c}=\sum_{n=0}^{\infty}\zeta^{-2n}c_{2n}.
    \end{array}
\end{equation}
Then we obtain from (\ref{eq27}) the mKdV hierarchy
\begin{equation}\label{eq29}
    q_{t_2}=-q_x;\quad
    q_{t_4}=\frac14(\pm 6q^2q_x- q_{xxx});\ldots .
\end{equation}
In a similar way, negative flows correspond to expansions in powers of $\zeta^2$:
\begin{equation}\label{eq30}
\begin{array}{l}
    \frac{\prt}{\prt\tau}=\sum_{n=1}^{\infty}{\zeta^{2n}}\frac{\prt}{\prt t_{-2n}},\\
    {a}=\sum_{n=1}^{\infty}\zeta^{2n}a_{-2n},\quad
    {b}=\sum_{n=1}^{\infty}\zeta^{2n}b_{-2n},\quad
    {c}=\sum_{n=1}^{\infty}\zeta^{2n}c_{-2n}.
    \end{array}
\end{equation}
Substitution of these expansions into (\ref{eq27}) yields the systems
\begin{equation}\label{eq31}
    qq_{t_{-2}}=\pm a_{-2,x},\quad \left(\frac{a_{-2,x}}{4q}\right)_x=\pm a_{-2}q;
\end{equation}
\begin{equation}\label{eq32}
\begin{array}{l}
    qq_{t_{-4}}=\pm a_{-4,x},\quad \left(\frac{a_{-2,x}}{4q}\right)_x=\pm a_{-2}q,\\
    \left(\frac14q_{xt_{-4}}-qa_{-4}\right)_x=
    \pm\frac{4a_{-2,x}}{q}; \ldots ,
    \end{array}
\end{equation}
and expansion of (\ref{eq26}) yields the corresponding $t_{-2n}$ systems:
\begin{equation}\label{eq33}
\left(
\begin{array}{c}
\psi_1 \\ \psi_2
\end{array}
\right)_{t_{-2}}=\frac1{\la}
\left(
\begin{array}{cc}
a_{-2} & \mp\frac{a_{-2,x}}{2q}\\
\frac{a_{-2,x}}{2q} & -a_{-2}
\end{array}\right)
\left( \begin{array}{c}
\psi_1 \\ \psi_2
\end{array}\right),
\end{equation}
\begin{equation}\label{eq34}
\left(
\begin{array}{c}
\psi_1 \\ \psi_2
\end{array}
\right)_{t_{-4}}=\frac1{\la^4}
\left(
\begin{array}{cc}
A_{-4} & B_{-4} \\
C_{-4} & -A_{-4}
\end{array}\right)
\left( \begin{array}{c}
\psi_1 \\ \psi_2
\end{array}\right),
\end{equation}
where
$$
\begin{array}{l}
A_{-4}= a_{-2}\la+a_{-4}\la^3,\\
B_{-4}=qa_{-4}\mp
\left(\frac{a_{-4,x}}{4q}\right)_x\mp\frac1{2q}(a_{-2,x}\la+a_{-4,x}\la^3),\\
C_{-4}=\pm qa_{-4}-
\left(\frac{a_{-4,x}}{4q}\right)_x+\frac1{2q}(a_{-2,x}\la+a_{-4,x}\la^3).
\end{array}
$$
It is easy to see that under proper normalization eqs.~(\ref{eq31}) reduce to sinh-Gordon or
sine-Gordon equation depending on the choice of the sign. For definiteness we choose the lower sign and
introduce 
\begin{equation}\label{eq36}
    q=\frac12u_x,\qquad a_{-2}=\frac12\cos u.
\end{equation}
Then eqs.~(\ref{eq31}) reduce to the sine-Gordon equation
\begin{equation}\label{eq37}
    u_{xt_{-2}}=2\sin u.
\end{equation}
In a similar way the second negative flow (\ref{eq32}) takes the form
\begin{equation}\label{eq37a}
    u_xu_{xt_{-4}}=-4a_{-4,x},\quad \left(
    \frac14u_{xxt_{-4}}-u_xa_{-4}\right)_x=2\sin u.
\end{equation}
Apparently, these equations has not been written down explicitly so far.

Note that generating equations (\ref{eq27}) can be written in compact form after
substitution of $q=\frac12u_x$ and exclusion of $a,\,b,\,c$ variables:
\begin{equation}\label{38}
    \left(\frac{u_{xx\tau}-4\zeta^2 u_{\tau}}{u_x}\right)_x=\pm u_xu_{x\tau}.
\end{equation}
It is easy to check that the first two terms of the expansion of this equation
(with choice of lower sign) yield equations
equivalent to (\ref{eq37}) and (\ref{eq37a}). The next higher negative flows can
be written down explicitly by this method without any difficulties.

At last, we remark that the above approach can be applied to derivation of
the generating function for the Whitham modulation equations in the AKNS hierarchy.
In a finite gap method of obtaining periodic solution of integrable equations,
a one-phase solution of equations belonging to the AKNS hierarchy is usually
parameterized by four zeros $\la_i,\,i=1,2,3,4,$ of the fourth degree polynomial
(see, e.g. \cite{Kamch2000})
\begin{equation}\label{eq39}
    P(\la) =\prod_{i=1}^4(\la-\la_i).
\end{equation}
The wavelength of the periodic solution is given by the expression
\begin{equation}\label{eq40}
    L=\frac12\oint\frac{d\mu}{\sqrt{P(\mu)}}.
\end{equation}
In a modulated periodic wave the parameters $\la_i$ become slow functions
of the space and time variables and their evolution is governed by the
Whitham modulation equations. Derivation of the generating function of these equation
actually coincides with the derivation of the Whitham equations for the SIT
system \cite{Kamch2000,KP95} and therefore we shall omit here the details.
The result for the generating function of the Whitham equations in the AKNS
hierarchy has the form
\begin{equation}\label{eq41}
    \frac{\prt \la_i}{\prt\tau}+v_i(\la)\frac{\prt \la_i}{\prt x}=0,
\end{equation}
where
\begin{equation}\label{eq42}
    v_i(\la)=\left(1-\frac{L}{\prt_iL}\prt_i\right)\frac1{\sqrt{P(\zeta)}},\quad
    \prt_i\equiv\frac{\prt}{\prt\la_i},\quad i=1,2,3,4.
\end{equation}
Introduction of time variables according to (\ref{eq4}) or (\ref{eq9}) and expansion
of (\ref{eq42}) in powers of $1/\zeta$ or $\zeta$, respectively, yields the
Whitham equations for $\la_i$ corresponding to each member of the hierarchy. The
particular cases of (\ref{eq42}) were discussed in \cite{Pavlov} (for the KdV equations
case, when $P(\la)$ has three zeros $\la_i,\,i=1,2,3$) and \cite{Kamch01} (for the
NLS equation case).

In conclusion, we have found generating functions for positive and negative flows
belonging to the AKNS hierarchy and some its reductions. This gives a simple method
of derivation of integrable equations, their Lax pairs and Whitham modulation equations.
Some known equations
usually not considered as members of the AKNS hierarchy find their natural
place within the AKNS scheme in our approach.
This observation permits one to combine known equations into new integrable equations
of physical importance.

\subsection*{Acknowledgements} AMK is grateful to J.F.~Gomes and A.H.~Zimerman for
interesting discussions of negative flows in the AKNS hierarchy.
This work was partially supported by RFBR (grants
\mbox{00--01--00210}, \mbox{00--01--00366}, \mbox{01--01--00696}).

\end{document}